\documentclass{ws-ijmpd}
\usepackage[super,compress]{cite}
\usepackage{url}
\usepackage{hyperref}
\usepackage{xcolor}

\newcommand{\beqa}{\begin{eqnarray}}
\newcommand{\eeqa}{\end{eqnarray}}

\begin{document}

\markboth{Y. Yang, Q. Li, J. Hao, and X. Li}
{Cosmological abundance of primordial black holes}

%
\catchline{}{}{}{}{}
%

\title{Cosmological abundance of primordial black holes in mixed dark matter scenarios incorporating Kaluza-Klein dark matter}

\author{Yupeng Yang, Qianyong Li, Jiali Hao}

\address{School of Physics and Physical Engineering, Qufu Normal University\\
Qufu, Shandong, 273165, China\\
ypyang@aliyun.com}

\author{Xiujuan Li}

\address{School of Cyber Science and Engineering, Qufu Normal University\\
Qufu, Shandong, 273165, China}

\maketitle

\begin{history}
\received{Day Month Year}
\revised{Day Month Year}
\end{history}

\begin{abstract}
The lightest Kaluza-Klein (KK) dark matter particles and primordial black holes (PBHs) emerge as plausible candidates for dark matter. In scenarios where dark matter is a mix of KK particles and PBHs, PBHs can attract surrounding KK dark matter particles post-formation, leading to the creation of ultracompact dark matter halos (UCMHs). The distribution of KK dark matter particles within UCMHs tends to be steeper than in classical dark matter halo structures, such as the Navarro-Frenk-White model. Consequently, the annihilation rate of KK dark matter particles in UCMHs is significant. The high-energy photons resulting from the annihilation of KK particles in UCMHs contribute to the extragalactic gamma-ray background (EGB). Leveraging data from the $\mathtt{Fermi\text{-}LAT}$  experiment, we have derived, for the first time, upper limits on the cosmological abundance of PBHs in the context of KK dark matter annihilation. For a KK dark matter mass range of $500\le m_{\rm B^{(1)}}\le 1500$ GeV, which aligns with the observed present abundance of dark matter, the conservative limits on the fraction of dark matter in PBHs, for massive PBHs with $M_{\rm PBH}\gtrsim 10^{-11}M_{\odot}$, are $f_{\rm PBH} \lesssim 2\times 10^{-5}$.
\end{abstract}

\keywords{dark matter; primordial black hole; Kaluza-Klein dark matter.}

\ccode{PACS numbers:95.35.+d}


\section{Introduction}	

Dark matter, a fundamental component of the Universe, has been confirmed through numerous astronomical observations. However, its nature remains an enigma to this day. Among the myriad dark matter models, the neutralino, predicted by supersymmetry theory, stands out as a popular candidate~\cite{2005PhR...405..279B,1996PhR...267..195J,Roszkowski:2004jc}. Despite extensive direct and indirect experimental searches, the neutralino has yet to be discovered (see, e.g., Refs.~\refcite{Feng:2013zca,Barger:2008su,PhysRevLett.118.191101,Ibarra:2013zia,XENON:2018voc,Arcadi:2017kky,PandaX-II:2016vec,DAMPE:2017fbg,Feng:2022rxt,Schumann:2019eaa}). 
An alternative dark matter candidate is the Kaluza-Klein (KK) dark matter particle, which is predicted by extra-dimensional theories~\cite{Appelquist:2000nn,Cheng:2002iz,Servant:2002aq}. The mass of the KK dark matter particle is dependent on the specific mode within the KK tower. Notably, the lightest KK particle (LKP), typically denoted as $B^{(1)}$, 
is stable and hence a potential dark matter candidate~\cite{Bergstrom:2004cy,Cheng:2002ej,Arrenberg:2008wy}. 
Fundamentally, the LKP falls under the weakly interacting massive particles (WIMPs) paradigm and shares similarities with the neutralino. 
Consequently, the direct and indirect search strategies employed for WIMPs are also applicable to the LKP. 
Despite this, numerous studies have focused on detecting the LKP. For instance, the authors of~\refcite{Cheng:2002ej,Barrau:2005au,Hooper:2005fj} 
have explored theoretical avenues for both direct and indirect detection of the LKP. These investigations encompass predictions 
of spin-(in)dependent proton cross sections, as well as potential signals such as positron excesses, antiproton fluxes, and neutrino fluxes. 
Moreover, it has been proposed that the LKP annihilation could account for the excess of galactic cosmic-ray 
electrons~\cite{2008Natur.456..362C,zhu2017gravitonmediated,Okada:2009bz}. Furthermore, the analysis of muon fluxes detected by the IceCube experiment 
can be utilized to impose constraints on the LKP dark matter annihilation occurring within the Sun~\cite{Abbasi_2010,ColomiBernadich:2019upo}.

Similar to the neutralino, KK particles can annihilate into fundamental particles~\cite{Bergstrom:2004cy,Cheng:2002ej,PhysRevD.68.044008,Tsuchida:2017guj,2009arXiv0906.3969D}.
To be consistent with the observed present abundance of dark matter ($\rm \Omega_{\rm DM}\sim 0.23$), 
the mass of the LKP is constrained to the range of $500\lesssim m_{\rm B^{(1)}}\lesssim 1500$ GeV~\cite{Appelquist:2000nn,2009arXiv0906.3969D,Bergstrom:2004cy}.

Primordial black holes (PBHs) represent another intriguing dark matter candidate, potentially formed through the direct collapse of large density perturbations 
(e.g., $\delta \rho/\rho > 0.3$) 
in the early universe~\cite{2021RPPh...84k6902C,Carr:2009jm}. In mixed dark matter scenarios comprising PBHs and KK dark matter, PBHs can attract surrounding KK dark matter particles to form ultracompact dark matter halos (UCMHs)~\cite{2009ApJ...707..979R}. 
These UCMHs exhibit a higher density of dark matter particles compared to standard dark matter halos, such as those described by the Navarro-Frenk-White 
model~\cite{Navarro:1996gj,Navarro:1995iw}. 
Given that the annihilation rate of KK dark matter particles is proportional to the square of their number density, the annihilation rate in UCMHs is substantial. Consequently, UCMHs are potential high-energy astrophysical sources. Previous research has primarily focused on the photon flux produced by neutralino annihilation in UCMHs~\cite{PhysRevD.85.125027,Scott:2009tu,2011JCAP...12..020Y,2023EPJC...83..934Y,2022PhRvD.105d3011Z}. 
In this work, we investigate for the first time the photon flux generated by KK dark matter particle annihilation in UCMHs, particularly its contribution to the extragalactic gamma-ray background (EGB). The EGB has been detected by the $\mathtt{Fermi\text{-}LAT}$ experiment~\cite{Fermi-LAT:2014ryh}, which has been utilized to study 
the relevant properties of dark matter, see, e.g., Refs.~\refcite{Yuan:2011yb,Liu:2016ngs,Shirasaki:2016kol,Fornasa:2011yb,Fermi-LAT:2010qeq}. 
Here we will use the EGB to investigate the cosmological abundance of PBHs in mixed dark matter scenarios. 

It should be noted that, in addition to WIMPs, other mixed dark matter scenarios involving PBHs have been 
extensively studied in the literature. For instance, the authors of Refs.~\refcite{Hertzberg:2019exb,Choi:2022btl,Nurmi:2021xds,Lasenby_2016} 
investigated hybrid dark matter models consisting of axions (or axion-like particles) and PBHs, where the minihalos around PBHs are composed of axion. 
Furthermore, the authors of Ref.~\refcite{Khlopov_2006} explored an alternative production mechanism for gravitino dark matter through 
the evaporation of PBHs.

This paper is organized as follows. In Sec.~\ref{sec2}, we briefly review the basic properties of UCMHs in the mixed dark matter scenarios consisting of PBHs and KK dark matter. 
In Sec.~\ref{sec3}, we derive the upper limits on the cosmological abundance of PBHs using the EGB data from the $\mathtt{Fermi\text{-}LAT}$ experiment, 
and then the conclusions are given in Sec.~\ref{sec4}. 

\section{The basic properties of UCMHs}
\label{sec2}

In the mixed dark matter scenarios consisting of cold particle dark matter and PBHs, UCMHs can be formed via the infall of particle 
dark matter onto PBHs~\footnote{{\bf Note that we do not delve into the specific mechanisms of their (co)genesis in the early universe here. 
Following previous studies~\cite{Eroshenko:2016yve,2021MNRAS.506.3648C,Boucenna:2017ghj,Boudaud:2021irr}, we assume that LKPs and PBHs 
are generated independently during the early universe's evolution. Apart from gravitational forces, these two components do not interact with each other.}}.
It is noteworthy that WIMPs are scarcely captured by PBHs prior to the kinetic decoupling epoch. 
This occurs because WIMPs remain tightly coupled to the primordial plasma until this epoch, thereby impeding their capture by PBHs~\cite{2021MNRAS.506.3648C}. 
Following previous works, such as Refs.~\refcite{2021MNRAS.506.3648C,Boudaud:2021irr,Gines:2022qzy,Boucenna:2017ghj}, 
we will subsequently examine the formation of UCMHs subsequent to the kinetic decoupling of Kaluza-Klein dark matter particles.

The density profile of particle dark matter within a UCMH varies depending on the 
masses of PBHs and particle dark matter. In general, for heavy PBH and particle dark matter, 
the density profile of particle dark matter within a UCMH is in a form of 
$\rho_{\rm DM}\propto r^{-3/2}$ and $r^{-9/4}$ from inside out~\cite{2009ApJ...707..979R}. 
For light PBH and particle dark matter, compared with the gravitational potential energy, the kinetic energy of 
particle dark matter has a significant effect on the density profile of UMCHs, resulting in a triple broken power law 
with $r^{-3/4}$, $r^{-3/2}$ and $r^{-9/4}$ from inside out. 
Specifically, the density profile of particle dark matter around a PBH with mass $M_{\rm PBH}$ 
can be written as~\cite{Eroshenko:2016yve,2021MNRAS.506.3648C,Boucenna:2017ghj,Boudaud:2021irr}, 

\beqa
\rho_{\rm DM}(r)=\left\{
\begin{array}{rcl}
&&f_{\rm DM}\rho_{\rm KD}\left(\frac{r_{c}}{r}\right)^{3/4}, ~~~~~~~~~~~~~~~~~~{r\leq r_{c}}\\
\\
&&f_{\rm DM}\frac{\rho_{\rm eq}}{2}\left(\frac{M_{\rm PBH}}{M_{\odot}}\right)^{3/2}\left(\frac{\hat r}{r}\right)^{3/2},~~{ r_{c}\leq r\leq r_k}\\
\\
&&f_{\rm DM}\frac{\rho_{\rm eq}}{2}\left(\frac{M_{\rm PBH}}{M_{\odot}}\right)^{3/4}\left(\frac{\bar r}{r}\right)^{9/4},~~~~~~~{ r >r_k}
\end{array} \right.
\label{eq:rho_r}
\eeqa
where $f_{\rm DM}\simeq 1$ is the fraction of dark matter in particle dark matter. $\hat r$ and $\bar r$ are written in the form of 

\beqa
{\hat r} = \frac{GM_{\odot}}{c^2}\frac{t_{\rm eq}}{t_{\rm KD}}\frac{m_{\rm DM}}{T_{\rm KD}}, ~~~{\bar r}=\left(2GM_{\odot}t^{2}_{\rm eq}\right)^{1/3},
\label{eq:r_ta}
\eeqa
where $r_c$ and $r_k$ can be obtained by the same density value at the junctions of different power laws,

\beqa
r_{c} = \frac{r_s}{2}\left(\frac{m_{\rm DM}}{T_{\rm KD}}\right), ~~~r_{k}=4\frac{(ct_{\rm KD})^2}{r_s}\left(\frac{T_{\rm KD}}{m_{\rm DM}}\right)^2,
\label{eq:r_rc}
\eeqa
where $r_s=GM/c^2$ is the Schwartschild radius. $t_{\rm KD}$ and $T_{\rm KD}$ stands for the time and temperature at 
kinetic decoupling and can be written as~\cite{Boucenna:2017ghj}, 

\beqa
t_{\rm KD} = \frac{2.4}{\sqrt{g_{\rm KD}}}\left(\frac{T_{\rm KD}}{\rm 1MeV}\right)^{-2}s, ~T_{\rm KD}=\frac{m_{\rm DM}}{\Gamma[3/4]}
\left(\frac{{\alpha}m_{\rm DM}}{M_{\rm Pl}}\right)^{1/4},
\label{eq:t_kd}
\eeqa
where ${\alpha}=\sqrt{16\pi^{3}g_{\rm KD}/45}$ with $g_{\rm KD}=61.75$. The cosmological density 
at the time of kinetic decoupling is $\rho_{\rm KD}={\Omega_m}\rho_{\rm eq}(t_{\rm eq}/t_{\rm KD})^{3/2}$~\cite{Eroshenko:2016yve}. 

For $r\to 0$, the density of particle dark matter goes to $\rho_{\rm DM} \to \infty$. On the other hand, the central density of a UCMH can be smoothed 
by the dark matter annihilation, resulting in a maximum density $\rho_{\rm max}$ at the center of UCMHs~\cite{Yang:2011ef,PhysRevD.72.103517,Hao:2024hzu}, 

\beqa
\rho_{\rm max} = \frac{m_{\rm B^{(1)}}}{\left<\sigma v\right>(t(z)-t_{i})},
\label{eq:rho_max}
\eeqa
where $t_{i}$ is the formation time of UCMHs, and ${m_{\rm B^{(1)}}}$ is the mass of LKP.  $\left<\sigma v\right>$ is the thermally averaged annihilation 
cross section and we adopt the form of~\cite{Servant:2002aq,Bergstrom:2004cy} 

\beqa
\left<\sigma v\right> =3\times10^{-26}\left(\frac{m_{\rm B^{(1)}}}{800\rm GeV}\right)^{-2}\rm cm^3s^{-1}
\label{eq:cross_section}
\eeqa

Including the dark matter annihilation, we set the density profile of particle dark matter within a UCMH as 

\beqa
\rho(r,z)={\rm min}\left[\rho_{\rm DM}(r), \rho_{\rm max}\right] 
\label{eq:rho_final}  
\eeqa

\begin{figure}
\centering
\includegraphics[width=0.9\textwidth]{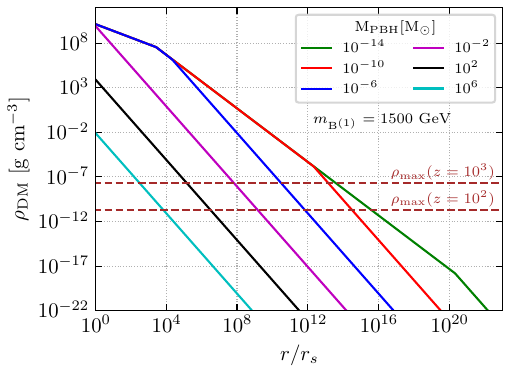}
\caption{The density profile of KK dark matter around different PBHs [Eq.~(\ref{eq:rho_r})]. 
Here we have set the KK dark matter mass $m_{\rm B^{(1)}}=1500$ GeV. 
The horizontal brown dashed lines show the maximum density $\rho_{\rm max}$ at redshift $z=100$ and 1000 [Eq.~(\ref{eq:rho_max})].
} 
\label{fig:rho_compare}
\end{figure}

For the case of neglecting the kinetic energy of particle dark matter compared to the potential energy, 
the mass of PBHs is approximately in the range  of 
$M_{\rm PBH} \gtrsim 10^{-9} M_{\odot }(m_{\rm B^{(1)}}/800\rm GeV)^{-19/8}$~\cite{2021PhRvD.103l3532T}. 
The density profile of KK dark matter particle within UCMHs are shown in Fig.~\ref{fig:rho_compare} for 
different PBH masses with $m_{\rm B^{(1)}}=1500$ GeV.

For the KK dark matter mass under consideration, the kinetic decoupling temperature can be estimated as $T_{\rm KD}\sim 100\rm MeV$, 
corresponding to a decoupling time of $t_{\rm KD}\sim 10^{-5} s$. 
In the standard formation scenario of PBHs, where PBHs are formed from the gravitational collapse of large density perturbations, 
the mass of a PBH formed at time $t$ is given by $M_{\rm PBH} \sim 10^{5}(t/1s)M_{\odot}$~\cite{Carr:2009jm}. 
Consequently, PBHs with masses of $M_{\rm PBH} > M_{\odot}$ should be formed after the kinetic decoupling epoch (i.e., at temperatures $T<100\rm MeV$). 
In this work, we will explore a broad mass range for PBHs, spanning $10^{-16}<M_{\rm PBH}<10^{4}~M_{\odot}$. Therefore, a significant fraction of low-mass 
PBHs ($M_{\rm PBH}<M_{\odot}$) should be formed before the kinetic decoupling of KK dark matter. 
As previously noted, although the formation of these low-mass PBHs before kinetic decoupling, 
KK particle dark matter is scarcely accreted onto PBHs to form ultracompact minihalos until kinetic decoupling. 

In addition to the standard scenario for the formation of PBHs, there exist numerous alternative formation mechanisms 
that diverge from the conventional one, see, e.g., Refs.~\refcite{2021RPPh...84k6902C,MaximKhlopov2010} for comprehensive review. 
In certain formation scenarios, such as those involving the collapse of cusps on cosmic strings~\cite{Jenkins:2020ctp} 
or occurring during the matter-dominated era of the Universe~\cite{Harada:2017fjm}, PBHs are generated much later compared to the standard scenario~\footnote{
Despite formed at a relatively late cosmic time, these black holes can still be designated as primordial black holes 
to differentiate them from astronomical black holes, which emerge from the gravitational collapse of massive stars.}, 
which posits their formation in the radiation-dominated phase of the Universe. Following the formation of these PBHs, WIMPs can also be accreted onto them to form minihalos. 
Nevertheless, the central density of the resulting dark matter halos is expected to be considerably lower than that observed in the standard scenario. 
This discrepancy can be partially attributed, for example, to the reduced number density of WIMPs in the surrounding environment compared to the standard scenario.


\section{The extragalatic gamma-ray flux from KK dark matter annihilation within UCMHs and constraints on PBHs}
\label{sec3}
Previous works have mostly focused on the extragalactic or galactic gamma-ray flux from neutralino dark matter annihilation in UCMHs~\cite{2011JCAP...12..020Y,2020EPJP..135..690Y,2022PhRvD.105d3011Z,PhysRevD.85.125027,Scott:2009tu,2021MNRAS.506.3648C}. 
Similar to previous works, the differential gamma-ray flux from UCMHs due to KK dark matter annihilation can be written as,

\begin{small}
\beqa
&&\frac{d\phi_{\rm \gamma}}{dE}= \frac{f_{\rm PBH}\Omega_{\rm DM}\rho_{\rm c,0}}{M_{\rm PBH}}\frac{c}{8\pi}\frac{\left<\sigma v\right>}{m_{\rm B^{(1)}}^2} \nonumber \\
&&\times\int_{0}^{z_{\rm up}}\,\frac{dz}{H(z)}\sum^{i}B_{i} \frac{dN_{\gamma}^{i}}{dE}(E^\prime,z)e^{-\tau(z,E)}\int\rho^2(r,z)dV,
\eeqa
\end{small}
where $H(z)=H_{0}\sqrt{{\rm \Omega_{DM}}(1+z)^{3}+{\rm \Omega_{\Lambda}}+{\rm \Omega_{\gamma}}(1+z)^4}$, 
$f_{\rm PBH}=\rm \Omega_{\rm PBH}/\Omega_{\rm DM}$ is the fraction of dark matter in PBHs, $E^\prime = E(1 + z)$ and $z_{\rm up} = m_{\rm B^{(1)}}/E-1$. $\tau(z,E)$ is 
the optical depth, which depends on the redshift and energy~\cite{Bergstrom:2001jj,Salamon:1997ac,Primack:2000xp}. $dN_{\rm \gamma}^{i}/{dE}$ is the energy spectrum of gamma-ray from KK dark matter annihilation with branching ratio $B_i$. Following previous works~\cite{Tsuchida:2017guj,Bergstrom:2004cy}, 
assuming the $5\%$ mass splitting at the first KK level, we consider the photon emission from decay or gragmentation of secondaries of KK dark matter annihilation, 
including 1) $20\%$ for each charged lepton; 
2) $11\%$ for each up-type quark; 3) $0.7\%$ for each down-type quark; 
4) $1\%$ for each charged gauge boson; 5) $0.5\%$ for each neutral gauge boson. We use the public code, e.g., $\mathtt{DarkSUSY}$~\cite{Bringmann:2018lay,Gondolo:2004sc}~\footnote{https://darksusy.hepforge.org/}, to calculate the energy spectrum. 

\begin{figure}
\centering
\includegraphics[width=0.9\textwidth]{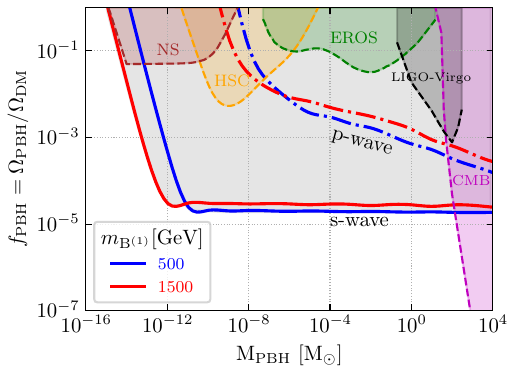}
\caption{Upper limits on the fraction of dark matter in PBHs in the mixed dark matter scenarios consisting of Kaluza-Klein dark matter 
and PBHs (blue and red solid lines). Constraints derived from several other measurements for comparison (dashed): 
1) the ultracompact binary search in advanced LIGO-Virgo (O2)(black)~\cite{LIGOScientific:2019kan,Kavanagh:2018ggo}; 
2) the anisotropy of cosmic microwave background (CMB) including the accreting PBHs in view of the Planck 2018 data (megenta)~\cite{2020PhRvR...2b3204S}; 
3) the gravitational lensing results measured by EROS (green) and HSC (orange)~\cite{EROS-2:2006ryy,Croon:2020ouk};
4) the disruption of neutron stars caused by their capture of PBHs (brown)~\cite{Capela:2013yf}. Data are taken from the zenodo~\cite{bradley_j_kavanagh_2019_3538999}. 
Note that the grey shaded areas with slashes are those excluded by extragalactic gamma-ray observations, 
which are applicable to all permitted mass ranges of KK dark matter. Additionally, the constraints for the p-wave annihilation scenario 
are illustrated (indicated by dot-dashed lines).} 
\label{fig:compare}
\end{figure}

The EGB has been observed by the $\mathtt{Fermi\text{-}LAT}$~\cite{Fermi-LAT:2014ryh} and $\mathtt{EGRET}$~\cite{EGRET:1997qcq,Strong:2004ry} 
experiments. Following the methods in~\refcite{Fermi-LAT:2010qeq}, 
we use the $\mathtt{Fermi\text{-}LAT}$ observations to get the conservative upper limits on the cosmological abundance of PBHs, 
which can be calculated as,

\beqa
\phi_{i}^{\gamma}\leq M_{i} + n\times \sigma_{i}
\eeqa
where $\phi_{i}^{\gamma}$ is the integrated flux from UCMHs due to KK dark matter annihilation in $i$th energy bin. 
$M_i$ and $\sigma_i$ are the measured flux and error 
of $i$th bin. $n=1.64$ corresponds to the $95\%$ confidence level. The final constrains on the fraction of dark matter 
in PBHs are shown in Fig.~\ref{fig:compare}. For massive PBHs with masses $M_{\rm PBH}\gtrsim 10^{-11}(10^{-12})M_{\odot}$, 
the kinetic energy of KK dark matter, with mass of $m_{B^{(1)}}=500(1500)~\rm GeV$, can be neglected compared with the gravitational potential energy, 
and the constraints does not nearly depend on the mass of PBHs, $f_{\rm PBH}\sim 2\times 10^{-5}(3\times 10^{-5})$. For light PBHs with masses 
$M_{\rm PBH}\lesssim 10^{-11}(10^{-12})M_{\odot}$, the constraints on $f_{\rm PBH}$ become less stringent as the PBHs mass decreases. 
The similar features can also be found for the neutralino case, see, e.g., 
Refs.~\refcite{PhysRevD.85.125027,Scott:2009tu,2011JCAP...12..020Y,2023EPJC...83..934Y,2022PhRvD.105d3011Z}.


For comparison, the constraints on $f_{\rm PBH}$ from several other studies are shown in Fig.~\ref{fig:compare}. 
Here we have set the mass range of PBHs as $M_{\rm PBH} \gtrsim 10^{-16} M_{\odot}$. For these massive PBHs, 
the constrains on $f_{\rm PBH}$ are mainly from investigations into gravitational lensing (EROS and HSC), gravitational waves (LIGO-Virgo) and radiation resulting 
from the accretion of baryonic matter (CMB), see, e.g., Refs.
~\refcite{LIGOScientific:2019kan,Kavanagh:2018ggo,2020PhRvR...2b3204S,EROS-2:2006ryy,Croon:2020ouk,Capela:2013yf,2022ApJ...928L..13Z}. 
For light PBHs, the constraints primarily come from the analysis of radiation resulting from the Hawking 
radiation effect, see, e.g., Refs.~\refcite{Su:2024hrp,Zhao:2024yus,Yang:2024pfb,Cang:2020aoo,Xie:2024eug,CDEX:2024xqm,Huang:2024xap,Boudaud:2018hqb}.
In the mixed dark matter scenarios consisting of neutralino and PBHs, the authors of~\refcite{2021MNRAS.506.3648C}\footnote{Note that the authors 
of~\refcite{2021MNRAS.506.3648C} have used a different method to get the upper limits. 
Specifically, they assessed the integrated flux, produced by the neutralino annihilation in UMCHs, in conjunction with $\mathtt{Fermi}$'s 
sensitivity (not the observed data) to derive the final limits on $f_{\rm PBH}$.} found that 
the constrains are $f_{\rm PBH} \lesssim 2\times 10^{-9}$ for $m_{\chi}=1\rm TeV$, 
$\left<\sigma v\right>_{\chi} =3\times10^{-26}~\rm cm^3s^{-1}$, which are not shown in Fig.~\ref{fig:compare}.

Note that astrophysical contributions are not included in this analysis, and we have derived conservative limits on the PBHs fraction $f_{\rm PBH}$. 
The extragalactic gamma-ray background is dominated by emissions from blazars~\cite{Ajello:2015mfa}, 
radio galaxies~\cite{Inoue_2011}, and star-forming galaxies~\cite{Ackermann_2012}. Specifically, for gamma-ray energies $E_{\gamma} >100~\rm MeV$, 
blazars contribute approximately 50\% of the EGB, while radio galaxies and star-forming galaxies account for about 10-30\%, respectively. 
In addition to the contribution from dark matter annihilation in UCMHs under consideration, other dark matter annihilation channels 
could also significantly contribute to the EGB. For instance, diffuse dark matter annihilation in the intergalactic 
medium~\cite{Fermi-LAT:2010qeq,10.1111/j.1365-2966.2010.16482.x}, annihilation within dark matter halos~\cite{Fermi-LAT:2010qeq,Yuan:2011yb,Liu:2016ngs,Hutten:2017cyu}, 
and annihilation in dark matter spikes around supermassive black holes~\cite{PhysRevD.89.043520} are all plausible sources. 
It is expected that if astrophysical contributions and these additional dark matter annihilation signals (e.g., from other dark matter structures) 
are accounted for, the derived constraints on $f_{\rm PBH}$ are expected to become more stringent~\cite{2022PhRvD.105d3011Z,Gines:2022qzy}.

In the preceding discussion, we assumed the thermally averaged annihilation 
cross section (Eq.~(\ref{eq:cross_section})) to be velocity independent, corresponding to the typical 
s-wave annihilation scenario. However, in general, the thermally averaged annihilation cross section 
$\left<\sigma v\right>$ can exhibit velocity dependence, as in the p-wave annihilation case. 
For p-wave annihilation, the cross section can be written as~\cite{Gines:2022qzy}:

\beqa
\left<\sigma v\right> = \left<\sigma v\right>_{0}^{p}\frac{v^2}{v_{0}^2}=\frac{\left<\sigma v\right>_{0}^{p}}{v_{0}^{2}}\frac{GM_{\rm PBH}}{r}
\eeqa
where $v_0$ is the velocity dispersion at freeze-out. For our calculations, we adopt $v_{0}^{2}=0.3$ and set 
$\left<\sigma v\right>_{0}^{p}=2\left<\sigma v\right>_{0}^{s}$~\cite{Kadota:2021jhg}, where $\left<\sigma v\right>_{0}^{s}$ is the 
the s-wave cross section (Eq.~(\ref{eq:cross_section})) at reference velocity. The resulting constraints on $f_{\rm PBH}$ are illustrated in Fig.~\ref{fig:compare} 
(dot-dashed lines). Overall, for the mass range of PBHs under consideration, the constraints derived for p-wave 
annihilation are weaker than those for s-wave annihilation.

Note that although both KK dark matter and neutralino belong to the category of WIMPs, and apart from their different origins, 
there are several distinct differences between them as follows:

$\bullet$  For KK dark matter, the thermally averaged annihilation cross section, as described by Eq.~(\ref{eq:cross_section}), 
is dependent on the mass of the KK dark matter, $m_{\rm B^{(1)}}$, whereas it is nearly independent of the mass for neutralino dark matter. 
Consequently, the constraints on $f_{\rm PBH}$, as shown in Fig.~\ref{fig:compare}, for KK dark matter depend solely on its mass. 
Furthermore, the energy spectrum of gamma-ray resulting from KK dark matter annihilation differs from that of neutralino annihilation. 

$\bullet$ Although the exact masses of both KK dark matter and neutralino remain unknown, the mass of KK dark matter has been constrained to 
a smaller range compared to that of the neutralino. As shown in Fig.~\ref{fig:compare}, 
the grey shaded areas with slashes are those excluded by gamma-ray observations, which are applicable to all permitted mass ranges of KK dark matter. 
In contrast, the neutralino has a much larger mass range and therefore exhibits greater uncertainty, see, e.g, Ref.~\refcite{2021MNRAS.506.3648C}.

\section{Conclusions}
\label{sec4}

In the mixed dark matter scenarios consisting of PBHs and KK dark matter particles, PBHs can accrete surrounding KK dark matter particles 
to form UCMHs subsequent to their formation. 
The number density of KK dark matter particles in UMCHs is larger than that of classical dark matter halos, leading to a high annihilation rate. 
It is anticipated that the KK dark matter annihilation in UCMHs have contributions to, e.g., the extragalactic gamma-ray background. 
Given that previous works have mainly focused on the neutralino dark matter, for the very first time, we here conducted a study on the KK dark matter 
and derived the upper limits on the cosmological abundance of PBHs using the EGB detected by the 
$\mathtt{Fermi\text{-}LAT}$ experiment. 
We considered the annihilation of the lightest KK dark matter, and found that 
for the lightest KK dark matter mass range $500 \lesssim m_{\rm B^{(1)}}\lesssim 1500$ GeV, which is allowed 
by the observed present abundance of dark matter, 
the conservative upper limit on the fraction of dark matter in PBHs is $f_{\rm PBH} \sim  10^{-5}$ 
for massive PBHs $M_{\rm PBH}\gtrsim 10^{-11}M_{\odot}$. For light PBHs, 
the constraints become less stringent as the PBHs mass decreases.

\section*{Acknowledgments}

We would like to thank Dr. Qiang Yuan for helpful comments. Yupeng Yang would like to express the gratitude to Feng Huang for her assistance in 
utilizing the public code $\mathtt{DarkSUSY}$. We are grateful to the anonymous referees for their instructive comments and suggestions.
This work is supported by the Shandong Provincial Natural Science Foundation 
(Grant Nos. ZR2021MA021 and ZR2023MA049).

\bibliographystyle{ws-ijmpd}
\bibliography{ref}

\end{document}